\documentclass{elsart}

\usepackage{amsmath}
\usepackage{amssymb}
\usepackage{graphicx}
\usepackage{dcolumn}
\usepackage{bm}

\def\be{\begin{eqnarray}}
\def\ee{\end{eqnarray}}
\def\ba{\begin{array}}
\def\ea{\end{array}}

\begin{document}

\begin{frontmatter}

\title{Investigations on unconventional aspects in the quantum Hall regime of narrow gate defined channels}

\author[l1]{J. Horas},
\ead{Jose.Horas@Physik.Uni-Muenchen.de}
\author[l1]{A. Siddiki},
\author[l1,l2]{J. Moser},
\author[l3]{W. Wegscheider} and
\author[l1]{S. Ludwig}

\address[l1]{Center for NanoScience and Dept. for Physics, Ludwigs-Maximillians University,
D-80539 Munich, Germany}
\address[l2]{
CNM-CSIC, Campus Universitat Autonoma de Barcelona, E-08193
Bellaterra, Spain}
\address[l3]{Institute for Experimental and Applied Physics, University of Regensburg,
D-93040 Regensburg, Germany}

\begin{abstract}
We report on theoretical and experimental investigations of the
integer quantized Hall effect in narrow channels at various
mobilities. The Hall bars are defined electrostatically in
two-dimensional electron systems by biasing metal gates on the
surfaces of GaAs/AlGaAs heterostructures. In the low mobility
regime the classical Hall resistance line is proportional to the
magnetic field as measured in the high temperature limit and cuts
through the center of each Hall plateau. For high mobility samples
we observe in linear response measurements, that this symmetry is
broken and the classical Hall line cuts the plateaus not at the
center but at higher magnetic fields near the edges of the
plateaus. These experimental results confirm the unconventional
predictions of a model for the quantum Hall effect taking into
account mutual screening of charge carriers within the Hall bar.
The theory is based on solving the Poisson and Schr\"odinger
equations in a self-consistent manner.
\end{abstract}
\end{frontmatter}
%

At low temperatures, transport in low-dimensional electron systems
is dominated by quantum mechanical properties. In particular, a
two dimensional electron system (2DES) subjected to a strong
magnetic field perpendicular to the 2DES exhibits the integer
quantized Hall effect (IQHE)~\cite{vKlitzing80:494}. In a
classical picture, with current driven in one direction ($x$), the
external magnetic field generates forward moving skipping orbits
along the edges of the sample due to the Lorentz force. The
quantization of this motion results in edge states, resembling
perfect one-dimensional channels, that explain basic features of
the IQHE~\cite{Girvin}. Twenty-five years after their discovery,
the standard explanation of the IQHE still employs a single
particle picture based on Landau quantization together with
localization of electronic states in a disordered background
potential. However, this picture discounts the Coulomb interaction
between electrons and, as a result, is unable to account for
several important features of the IQHE. These features include the
enormous accuracy with which the quantized resistance values can
be reproduced experimentally~\cite{Bachmair03:14} and the local
potential and current distributions in narrow Hall bars as
recently detected in "local probe"
experiments~\cite{Ahlswede01:562,Yacoby04:328}.

Chklovskii et al. first considered realistic electron-electron
interactions together with the confinement potential of narrow
Hall bars~\cite{Chklovskii92:4026,Oh97:13519}. In this theory,
strips of incompressible 2DES replace the edge states. A more
recent model takes interactions between electrons explicitly into
account within a self-consistent mean field
approximation~\cite{siddiki2004,Romer07}. Numerical calculations
within this model reproduce the main results of the "local probe"
experiments~\cite{siddiki2004}. In low mobility samples the
straight line, describing the magnetic field dependence of the
classical Hall resistance, cuts the Hall plateaus at its centers,
as also assumed within the standard explanations of the IQHE. In
contrast, the recent model predicts for even filling factors that
the classical Hall resistance line cuts the Hall plateaus at
higher magnetic fields in a high mobility narrow Hall bar. This
asymmetry is caused by the magnetic field dependence of the shape
of incompressible strips (ISs) separating compressible regions of
the 2DES~\cite{Siddiki:ijmp}. We find, that in high mobility
narrow Hall bars electron-electron interactions and the
confinement potential of the channels at the edges cause the
spatial distribution of ISs, which determine widths and positions
of the quantized Hall (QH) plateaus. Here, we briefly summarize
the findings of the model and compare them with experimental
results measured on high mobility samples with narrow
electrostatically defined channels.

\begin{figure}
\includegraphics[width=1.1\linewidth]{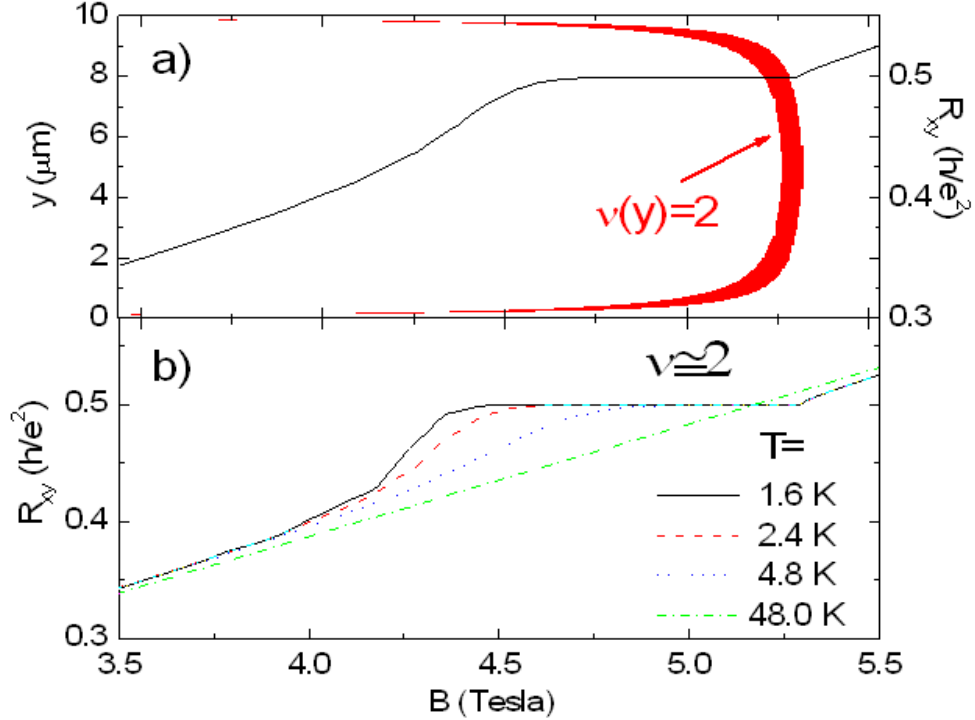}
{\centering \caption{ \label{fig:fig1} (a) Calculated position of
the incompressible strip at $\nu=2$ in a 10\,$\mu$m wide Hall bar
as a function of the magnetic field. For the calculation we used
$k_BT/E_F^0=0.02$ and assumed an average electron density of
$3.79\times10^{11}$ cm$^{-2}$. Also shown is the Hall resistance
$R_{xy}$ (black line) calculated for the same conditions. (b)
Calculated $R_{xy}$ as a function of magnetic field, at various
temperatures. }}
\end{figure}

The screening theory of the IQHE is based on the self-consistent
solution of the Poisson and Schr\"odinger equations to obtain the
electron density and electrostatic potential distribution within
the sample~\cite{siddiki2004}. The current distribution is
calculated by using a local version of Ohm's
law~\cite{Guven03:115327} incorporating a relevant model of
conductivity, e.\,g.\ self-consistent Born
approximation~\cite{Ando82:437}. Our numerical calculations start
at zero temperature and the electron density distribution at zero
field potential. The regions of incompressible 2DES and
compressible 2DES, respectively, at finite field and temperature
are then found in a self-consistent iterative calculation. The
numerical results in Fig.~\ref{fig:fig1}a
demonstrate the formation of an incompressible strip shown as a
function of the magnetic field strength across a narrow Hall bar.
Within this strip the local filling factor is exactly $\nu=2$
(note that the filling factor changes across the width of a narrow
Hall bar). Within the compressible regions adjacent to
incompressible strips the electrostatic potential is constant,
because of perfect screening and current only flows within
incompressible strips. Here, all current carrying states are
occupied. Thus, the longitudinal resistance $R_{xx}$ vanishes and
the Hall resistance $R_{xy}$ resumes its quantized plateau value
as also shown in Fig.~\ref{fig:fig1}a. Assuming spin degeneracy
for $\nu<2$ (high field side of the incompressible strip in
Fig.~\ref{fig:fig1}a)  only the lowest Landau level is partly
occupied and now incompressible strips are present. As a result of
the finite extend of the quantum mechanical wave functions of the
electrons in compressible regions very narrow incompressible
strips practically vanish~\cite{siddiki2004,Suzuki93:2986}. Hence,
for some magnetic field ranges at $\nu>2$ no ISs exist along the
Hall bar (compare Fig.~\ref{fig:fig1}a). At such magnetic fields
without incompressible strips $R_{xx}$ becomes finite and $R_{xy}$
deviates from its quantized value. Previous
approaches~\cite{Chklovskii92:4026,Oh97:13519} employing
Thomas-Fermi type approximations ignore the finite extent of the
wave functions and the number of incompressible strips is given by
the integer multiples of $\nu/2$ (assuming spin degeneracy).
Fig.~\ref{fig:fig1}b shows the calculated results for $R_{xy}$ in
the vicinity of the plateau at $\nu=2$ for four different
temperatures. Temperature broadening first destroys the
incompressible strip starting from its narrow side. Hence, with
increasing temperature deviations from the plateau happen at the
low field side and the highest temperature curve cuts through the
plateau nearly at its high field edge. Note that in the case of
low mobility, disorder widens the incompressible strips in respect
to the magnetic field and causes the plateau to extend to higher
magnetic fields.

\begin{figure}
{\centering
\includegraphics[angle=-90,width=1.\linewidth]{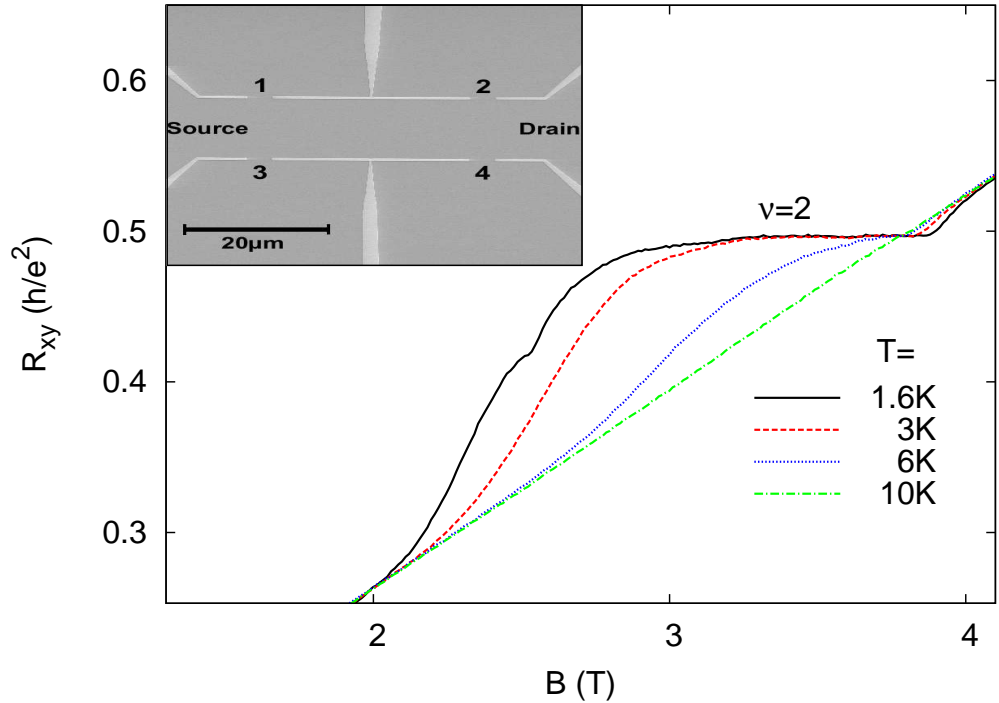}
\caption{ \label{fig:fig2} Measured Hall resistance $R_{xy}$ of a
gate defined high mobility Hall bar for different temperatures in
the vicinity of $\nu=2$. The inset shows a scanning electron
micrograph of the samples surface. Light gray areas are metal
gates used to define the Hall bar with a width of 10\,$\mu$m.}}
\end{figure}

We conducted experiments on several samples with different
geometries. They were all performed on GaAs/AlGaAs
heterostructures containing a 2DES 100\,nm below the surface. The
electron density and mobility of the 2DES used for the
measurements shown in Fig.~\ref{fig:fig2}
are $n_{\rm e}=1.8\times10^{11}$\,cm$^{-2}$ and
$\mu=2.96\times10^6$\,cm$^{2}/$Vs. The inset of
Fig.~\ref{fig:fig2} displays a scanning-electron-microscope
picture of the sample surface. All metallic gates (lighter gray)
are biased with $V_{\rm g}=-0.25$ V in respect to the 2DES in
order to define a 10\,$\mu$m wide Hall bar. At this gate voltage
the 2DES beneath the gates is locally depleted, while the contacts
(openings between gates) are still open. Instead of etching a Hall
bar we use gated structures in order to provide a lateral
confinement potential as smooth as possible. A small source-drain
AC-current is applied, meanwhile the Hall resistance  $R_{xy}$ is
measured using the contacts 1-3 (or 2-4) and $R_{xx}$ is measured
using 1-2 (or 3-4).

In Fig.~\ref{fig:fig2} we show $R_{xy}$ for several temperatures
measured in the vicinity of filling factor $\nu=2$. Comparison
with the numerical calculations (Fig.~\ref{fig:fig1}b) shows
excellent qualitative agreement. For a similar measurement on a
sample with a smaller mobility of $\mu\sim
1\times10^6$\,cm$^{2}/$Vs and an identical 10\,$\mu$m wide Hall
bar we find the expected low mobility behavior, namely that the
high temperature line of $R_{\rm xy}$ cuts through the center of
the quantized plateau.

In summary, on a gate defined Hall bar in a high mobility 2DES we
observe an asymmetric contraction of the plateaus of the Hall
resistance as temperature increases. The plateaus mainly shrink on
their low magnetic field side finally causing the high temperature
curve of the classical Hall resistance to cut through the plateau
at its high magnetic field side. For low mobility samples we
observe the usual symmetric behavior, where the high temperature
curve cuts through the center of the plateau. Our findings
perfectly match the predictions of a model taking into account
electrostatic interactions of electrons and the overlap of their
quantum mechanical wave functions.

\end{document}